# Aligning LLM-Simulated and Human Examinees for Psychometric Calibration: A Cognitive Diagnostic Profiling Approach*

Wenjie Zhou[1†] Yunting Liu[1] Renjiao Tang[2] Mark Wilson[1]

[1] Berkeley School of Education, University of California, Berkeley
[2] Department of Family Social Science, University of Minnesota Twin Cities, Minneapolis

## Abstract

Psychometric calibration for educational tests typically requires costly human response data. Large language models (LLMs) simulated examinees offer a promising route to early calibration, but their responses are too accurate and too uniform. We propose Cognitive Diagnostic Profiling (CDP), a zero-shot framework that prompts LLMs to simulate plausible examinees with diverse cognitive profiles: binary attribute-mastery patterns are rendered as natural-language profiles and sampled under an uninformative or an informative distribution. Using the Tatsuoka fraction-subtraction dataset (536 examinees, 15 items, five attributes), we evaluated eight LLM configurations under no-profile, uninformative-CDP, and informative-CDP conditions, assessing alignment with human examinees at the ability-distribution, mastery-profile, and item-difficulty levels. CDP improved all three levels: distributional overlap rose across configurations; weighted correlations between profile-level scores and human profile expectations reached 0.92 to 0.98; and item-difficulty recovery improved in rank order and absolute alignment, most for reasoning-enabled models; in the strongest case, Gemini 3.0 Flash (Thinking), one-parameter logistic (1PL) difficulty Spearman correlations rose from 0.24 to 0.86 and 0.90 and the root-mean-square error (RMSE) fell from 6.31 to 1.30 and 0.90; the informative condition helped most where profile-level alignment was strong. CDP brings LLM-simulated examinees into closer psychometric alignment with human examinees, making them practical for operational test development.

† Corresponding author: Wenjie Zhou. Email: zhouwenjie@berkeley.edu
* The data this study generated and used are openly available on https://osf.io/p6fam/overview?view_only=d149f9031cc645748c230f1d62f190be.

# 1 Introduction

Psychometric calibration, the estimation of an assessment's item and person parameters from response data, is a standard step in test development. Before items can be placed into operational use, their psychometric properties, primarily item difficulty, must be evaluated empirically. Because every response is the product of an examinee interacting with an item, this process characterizes not only the items but also the examinees who answer them, and it yields dependable item parameters only when the response data come from a sample that is representative of the target population and large enough for stable estimation (Haladyna & Rodriguez, 2013). Collecting such data is costly and time-consuming, requiring sample recruitment, standardized administration, and subsequent analysis. The costs become substantial in contexts where new items must be produced frequently. Adaptive testing, personalized learning, and online course platforms all rely on continually refreshed item pools, so pretesting every new item on a representative human sample imposes a real burden on development, creating practical value in methods that estimate item parameters at lower cost.

Several alternatives have been explored. Expert judges can rate item difficulty directly, though their estimates tend to agree only moderately with empirical values (Impara & Plake, 1998). Explanatory item response models, such as the linear logistic test model, estimate the contribution of each cognitive operation to item difficulty from existing items and use these estimates to predict difficulty for new items that draw on the same operations (Fischer, 1973; De Boeck & Wilson, 2004). Item-feature-based difficulty prediction uses natural language processing to relate linguistic and structural properties of items to their empirical difficulty, with promising results in several content areas (Ha et al., 2019; AlKhuzaey et al., 2024; Peters et al., 2025). These methods provide difficulty estimates for individual items, but leave open the question of whether complete examinee response data can be simulated for psychometric calibration.

## 1.1 LLMs as Simulated Examinees

Large language models (LLMs) offer a more direct pathway. If an LLM can generate item responses that resemble those of human examinees, item parameters can be estimated from LLM-generated data rather than from human pilot samples. The rationale goes beyond speed and cost: LLMs are trained on large corpora of human-generated text and show reasoning patterns that parallel those of human reasoners, including content effects on deductive tasks (Lampinen et al., 2024), and features of the LLM's internalized problem-solving process, such as solution-step count, cognitive complexity, and candidate misconceptions, predict empirical item response theory (IRT) difficulty on unseen mathematics items (Hoyl, 2026). Used directly as zero-shot respondents, LLMs reproduce human item statistics unevenly: they can preserve much of the relative ordering of item difficulty, but the extent of this preservation varies widely across content domains and tests (Liu et al., 2025; Säuberli et al., 2025). A useful simulation must both preserve the relative ordering of items and align with the target distribution in absolute terms; the evidence shows a clear split between these conditions, which the literature attributes to two mechanisms.

***Less alignment.*** This split is visible within a single study: GPT-3.5 recovered the relative ordering of 1PL difficulty well (Pearson 0.83, Spearman 0.87; Liu et al., 2025). Absolute alignment fails on the same example: the corresponding RMSE was 1.90 on items whose difficulties span only −2 to 2, an error large enough to undermine practical use. More broadly, LLM pass rates and parameter values diverge from human values in systematic ways, a failure we call less alignment that reflects a training-induced location bias. On a 1,031-item national high-stakes achievement test in Brazil, for example, ten LLMs systematically underestimated item difficulty (Brant et al., 2026). Because LLMs are optimized to produce correct, high-quality outputs, strong general-purpose models respond as far stronger examinees than the human population, reaching the 95th to 99th percentile of the human distribution on subject-by-grade combinations of National Assessment of Educational Progress (NAEP) mathematics and reading (Srivatsa et al., 2025), and most maintain near-ceiling accuracy under zero-shot persona prompting, with OpenAI's GPT o1 holding 99.0% across personas (Senthil Kumar et al., 2025). Wu et al. (2025) put the conflict directly: LLMs, "typically trained as 'helpful assistants', target at generating perfect responses" and "struggle to simulate students with diverse cognitive abilities." The bias can even worsen as models improve: Benedetto et al. (2024) described a "curse of hyper-accuracy" in which GPT-4, despite stronger reasoning, scored lower than GPT-3.5 on a response-monotonicity score, which indexes whether responses grow more accurate as the prompted ability rises, for the AI2 Reasoning Challenge (ARC) science benchmark (0.74 versus 0.86). He-Yueya et al. (2024) likewise found significant model–human misalignment and, surprisingly, that smaller models sometimes align better than larger ones.

***Less variability.*** The second mechanism concerns the spread of LLM responses. Even when prompts shift LLM behavior, the resulting distributions cluster within a narrow band: the 18 open-weight LLMs benchmarked by Säuberli et al. (2025) converged toward one another rather than toward the human distribution, and across nine open and commercial models, response distributions likewise failed to reproduce the variation of human survey respondents (Tjuatja et al., 2024). When the simulated population lacks spread, item-difficulty estimates become unreliable, especially for easy items, where nearly all simulated examinees respond correctly regardless of item properties; the compressed spread also shrinks the scale of the estimated difficulties, attenuating their range relative to the human calibration. Together, less alignment biases the location of LLM-derived parameters and less variability compresses their spread; both block absolute alignment, while relative structure can still partially emerge because LLMs distinguish easier from harder items within their narrow range.

## 1.2 Profile-Based LLM Simulation as a New Direction

Both problems point to the same need: LLMs require explicit, structured guidance to accurately simulate the diverse range of examinees in a real population, particularly those with partial understanding of the assessed construct. This has led to approaches that provide the LLM with an explicit profile of the simulated examinee (Tian et al., 2025; Marquez-Carpintero et al., 2025). Three main methods have emerged: multi-LLM aggregation, which combines response distributions from several models to enhance diversity (Liu et al., 2025) but lacks a principled mixture framework; fine-tuning, which adjusts a model to a specific population through supervised or preference-based training (Scarlatos et al., 2025; Maeda, 2025; He-Yueya et al., 2024; Ormerod, 2026), though it is computationally intensive and susceptible to the Student Data Paradox, where training on examinee errors enhances misconception simulation but reduces general reasoning ability (Sonkar et al., 2024; Tian et al., 2025); and prompt-based profiling, which embeds the examinee's background, ability, or skills directly into the prompt (Srivatsa et al., 2025; Senthil Kumar et al., 2025; Nguyen et al., 2025; Wu et al., 2025; Li et al., 2025; Acquaye et al., 2026). Among these, prompt-based profiling provides the best balance of cost and adaptability, requiring no training and adapting to new tests through prompt changes alone.

Existing prompt-based strategies have explored several ways to steer LLM behavior toward different ability levels, with varying degrees of success. Prompting with a numeric ability value shows limited effect zero-shot ($\theta$-alignment near 0.02 to 0.04), though alignment improves substantially under dedicated fine-tuning (0.73 to 0.77 with the SMART pipeline; Scarlatos et al., 2025). Persona or demographic prompts have shown limited effects on item-level performance (Senthil Kumar et al., 2025; Brant et al., 2026). Standardized proficiency-level role play has been more effective: simulating the four NAEP achievement levels in population proportions recovered the rank order of NAEP mathematics difficulties with correlations of 0.75 to 0.82 (Acquaye et al., 2026), though the descriptors operate along a single dimension. These approaches also differ in how they represent examinee ability. A numeric target such as "respond as a student with $\theta = -1.5$" (e.g., Scarlatos et al., 2025) relies on a psychometric scale that the LLM was not trained to interpret. A vague label such as "a below-average student" (e.g., Benedetto et al., 2024; Srivatsa et al., 2025; Senthil Kumar et al., 2025) conveys a general level without detailing the nature of the examinee's understanding. These descriptors place each examinee on a single ability continuum, but examinees at the same overall level can hold qualitatively different knowledge structures, a source of variation that unidimensional cues are not designed to express. These observations motivate the framework we propose next.

## 2 The Present Study: Cognitive Diagnostic Profiling

Building on recent work using LLMs as simulated examinees and on evidence that profile-based prompting can improve alignment with human performance, this study proposes Cognitive Diagnostic Profiling (CDP). CDP draws on cognitive diagnostic models (CDMs) and related discrete latent-class models (de la Torre, 2009, 2011; Tatsuoka, 2002; von Davier, 2008) to represent each simulated examinee as a binary attribute-mastery profile and prompts an LLM to generate item responses consistent with that profile. In CDMs, examinees are classified into mastery patterns, binary vectors indicating which cognitive skills each examinee has mastered, and item responses are modeled as functions of these patterns. CDP adopts the same representational structure but applies it generatively: mastery patterns are assigned to simulated examinees, rendered into natural-language skill statements, and supplied to an LLM, which generates responses conditioned on the stated profile. The resulting response matrix is then submitted to standard psychometric analysis.

CDP takes as input a set of design choices about the test and the target population and produces a prompt-level specification of each simulated examinee's cognitive profile in natural language. We describe it as six sequential steps (Figure 1): Steps 1 to 3 translate the construct into a natural-language profile, Step 4 assembles the examinee pool by sampling profiles under an uninformative or informative distribution, and Steps 5 to 6 generate the responses and submit the matrix to psychometric analysis.

**Figure 1**

*The six-step Cognitive Diagnostic Profiling (CDP) framework*

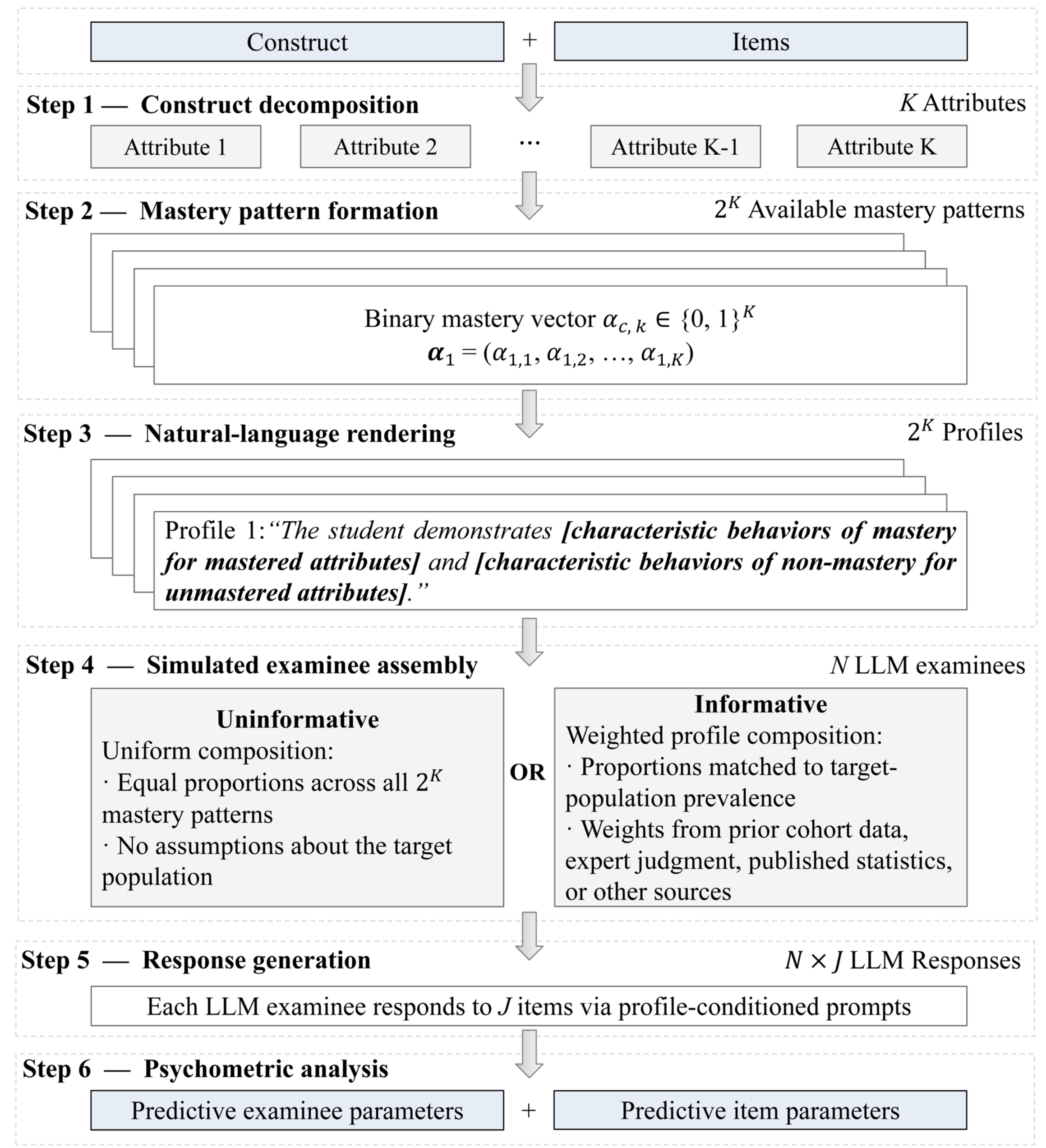

Step 1: Construct decomposition. The construct that the test measures is decomposed into $K$ discrete binary attributes: Attribute 1, Attribute 2, …, Attribute K. Each corresponds to a specific cognitive skill that the test aims to assess. This decomposition is standard in CDM practice and is typically established through content analysis by subject-matter experts (e.g., Akbay et al., 2017), sometimes validated against empirical item-response patterns (de la Torre, 2008).

Step 2: Mastery pattern formation. Each simulated examinee's cognitive profile is represented by a binary vector $\boldsymbol{\alpha} = (\alpha_1, \alpha_2, \ldots, \alpha_K)$ with $\alpha_k \in \{0, 1\}$, indicating which attributes the examinee has mastered ($\alpha_k = 1$) or not mastered ($\alpha_k = 0$). With $K$ binary attributes, there are $2^K$ possible mastery patterns, each representing a qualitatively distinct type of examinee. For $K = 5$, this yields 32 patterns ranging from $\alpha = (0, 0, 0, 0, 0)$, an examinee who has mastered no attribute, to $\alpha = (1, 1, 1, 1, 1)$, an examinee who has mastered all.

Step 3: Natural-language rendering. For a given pattern $\alpha$, each attribute value is translated into a natural-language skill statement. A mastered attribute ($\alpha_k = 1$) becomes a positive statement describing what the simulated examinee can do reliably for that skill, and an unmastered attribute ($\alpha_k = 0$) becomes a negative statement noting the absence of that skill. The K statements together constitute the prompt-level profile that is later supplied to the LLM.

Step 4: Simulated pool assembly. The sampling distribution over the $2^K$ patterns determines which profiles are drawn and in what proportions. We distinguish two cases by whether an information source about the target population is available:

Without an information source. When no target population information is used, each of the $2^K$ patterns is sampled with equal probability. This uniform allocation is designed to spread simulated examinees across the full range of mastery patterns and hence across the ability continuum; the realized ability distribution depends on the generating LLM.

With an information source. Patterns are sampled in proportion to their prevalence in the target population, with the weighting informed by any available source: a prior CDM fit on a comparable cohort, expert judgment, or published population statistics. The pattern proportions can be specified at three levels of granularity (attribute marginals, attribute pairs, or the full joint distribution).

Step 5: Response generation. Each examinee in the simulated pool is presented with the $J$ test items as natural-language prompts conditioned on its assigned profile, and the LLM's graded responses populate an $N \times J$ response matrix $Y$.

Step 6: Psychometric analysis. Standard psychometric models (e.g., CTT, IRT) are then applied to $Y$ to estimate examinee and item parameters that can be compared against human-examinee references.

Across the six steps, the framework separates three design decisions that are often conflated in prompt engineering: what the test measures (Step 1), how an examinee's cognitive state is represented (Steps 2–3), and how the simulated population is distributed (Step 4). Steps 1–3 and the uninformative Step 4 require only an attribute-level description of the test; the informative variant adds any available population information. The framework operates zero-shot, requires no fine-tuning, and can, in principle, be applied to any assessment for which such attribute descriptions can be written.

The framework follows assessment design traditions that ground measurement in an explicit model of cognition, a principle summarized in the assessment triangle (National Research Council, 2001). Evidence-Centered Design builds student models from cognitive claims that link observed behavior to underlying knowledge (Mislevy et al., 2003). Construct mapping defines proficiency levels by the skills examinees demonstrate at each stage (Wilson, 2023). Closest to CDP, the cognitive design system generates test items from a cognitive model (Embretson, 1998). CDP applies the same generative logic in the opposite direction: it generates examinees rather than items, translating each attribute pattern into natural-language skill statements that an LLM can interpret and act on. Prior work supports the feasibility of this approach: LLMs have been used to build and validate Q-matrices for TIMSS mathematics assessments (F1 ≈ 0.83; Sachdeva & Park, 2025), to draft and refine Q-matrices for diagnostic tests (Du et al., 2026), and profiles based on knowledge components derived from prompts have improved item evaluation accuracy ($r$ = .72; Lu & Wang, 2024).

The present study evaluates the effectiveness of the CDP framework: whether profiles generated through CDP, rendered into natural-language prompts, and supplied to LLMs yield simulated response data that usefully approximate human-examinee data. We compare a no-profile baseline against an uninformative profile condition and an informative profile condition. Three research questions guide this investigation:

1. Distribution-level alignment. Does CDP improve the alignment between the overall ability distribution of LLM-simulated examinees and that of human examinees?

2. Profile-level alignment. Beyond the overall distribution, do simulated examinees assigned a given mastery profile attain the ability and score expected of human examinees holding that profile?

3. Item-level recovery. Does CDP improve the recovery of item difficulty in both relative ordering and absolute alignment?

# 3 Methods

## 3.1 Dataset

Figure 2 provides an overview of the study design: panel (a) shows the human reference data described in this section, panel (b) the experiment design crossing three prompting conditions with eight LLM configurations (Section 3.2), and panel (c) the three-level evaluation framework (Section 3.3).

**Figure 2**

*Overview of the study design: human reference data, LLM simulation, and three-level evaluation*

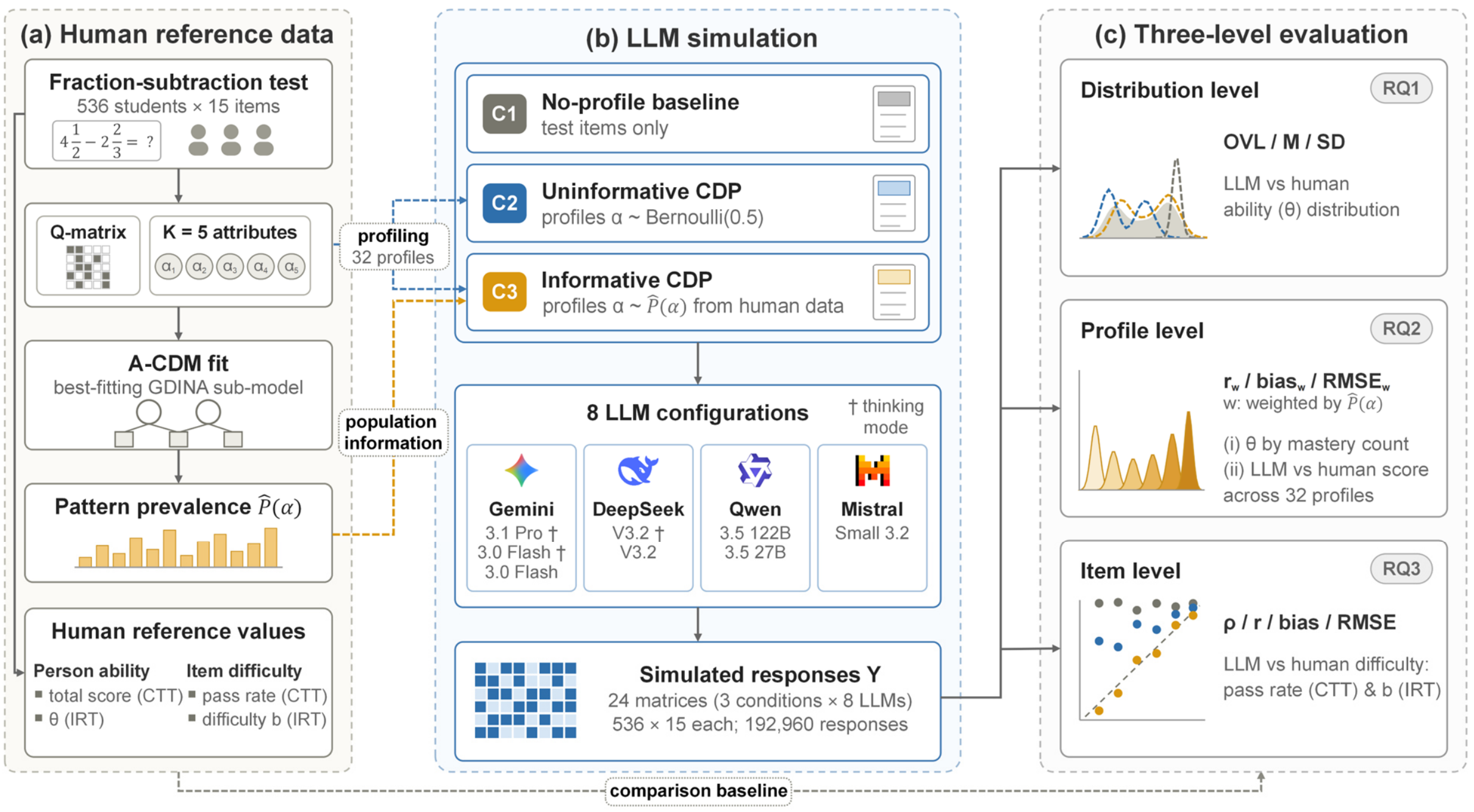

*Note.* LLM = large language model; CDP = cognitive diagnostic profiling; A-CDM = additive cognitive diagnosis model; GDINA = generalized deterministic-inputs, noisy-and-gate model; $\hat{P}(\alpha)$ = estimated mastery-pattern (profile) prevalence; OVL = overlapping coefficient; *M* = mean; *SD* = standard deviation; $\rho$ = Spearman rank correlation; *r* = Pearson correlation; RMSE = root-mean-square error.

We use the fraction-subtraction data originally collected by (Tatsuoka, 1990), with N = 536 middle-school students and the 15-item, five-attribute test form from de la Torre (2009), as provided in the CDM R package (George et al., 2016). The 15 items probe mixed-fraction subtraction and were designed to assess five binary cognitive attributes: $\alpha_1$ basic fraction subtraction, $\alpha_2$ simplifying and reducing fractions, $\alpha_3$ separating whole number from fraction, $\alpha_4$ borrowing from whole number to fraction, and $\alpha_5$ converting whole number to fraction. Full attribute definitions and the natural-language skill statements used in the LLM prompt are given in Appendix A.

Throughout the analysis, the human-examinee data serve as the target reference: per-item pass rates and 1PL item-difficulty estimates from the 536 human responses define the empirical reference against which LLM-simulated responses are compared. The dataset's Q-matrix (de la Torre, 2009) is used with the human responses to fit a CDM; we compare six sub-models of the generalized DINA (deterministic inputs, noisy "and" gate) framework (GDINA; de la Torre, 2011) and select the best-fitting one, the logistic A-CDM (an additive CDM with a logit link), which shows best fit across these six sub models. The posterior pattern prevalence $\hat{P}(\alpha)$ from this fit additionally anchors the C3 condition.

## 3.2 Experimental Design

We evaluate eight LLM configurations from three vendors (Figure 2b): Gemini 3.1 Pro (Thinking), Gemini 3.0 Flash (Thinking), Gemini 3.0 Flash (Non-Thinking), DeepSeek V3.2 (Thinking), DeepSeek V3.2 (Non-Thinking), Qwen 3.5 122B, Qwen 3.5 27B, and Mistral Small 3.2. The Gemini configurations are queried via the Google Generative Language API; the DeepSeek and Mistral configurations via OpenRouter; and the two Qwen configurations, the 122B mixture-of-experts (MoE) and 27B variants, via DashScope (Alibaba Cloud). Thinking mode is enabled for Gemini 3.1 Pro, one Gemini 3.0 Flash configuration, and one DeepSeek V3.2 configuration, as their names indicate, and disabled for all others. All API calls use each LLM's default sampling temperature (1.0). Responses, returned as a JSON object, include answers graded by exact match against a reference. Complete prompt templates are provided in Appendix A.

In every condition, each of the 536 simulated examinees completes the full 15-item test in a single response, yielding 536 × 15 = 8,040 item responses per cell, and 8 × 3 × 8,040 = 192,960 responses across the full design.

Condition 1 (C1): No-profile baseline. The LLM receives only the test items and a short instruction to predict a typical examinee's answers. C1 characterizes what the LLM produces when it receives no guidance about the examinee's cognitive state.

Condition 2 (C2): Uninformative CDP. Each of the 536 simulated examinees is assigned a mastery pattern $\alpha$ with no use of any information about the target population: we independently draw each $\alpha_k \sim Bernoulli(0.5)$, equivalent to sampling uniformly over the $2^5 = 32$ patterns. The pattern is rendered into a natural-language profile (Step 3) and prepended to the items.

Condition 3 (C3): Informative CDP. As in C2, each simulated examinee is assigned a rendered mastery pattern $\alpha$, now drawn from a categorical distribution with pattern probabilities $\hat{P}(\alpha)$ , the posterior pattern prevalence of the best-fitting CDM on the human-examinee data (the full joint-distribution granularity of Step 4). The mastery information for the target student population thus serves as the source of information.

## 3.3 Psychometric Analysis

We characterize item properties using two difficulty indices: the classical pass rate from classical test theory (CTT), defined as the proportion of examinees who answer an item correctly, and the item-difficulty parameter from IRT. On the person side, CTT summarizes each examinee by the total score, and the IRT fit yields the ability estimate. Throughout this paper, the IRT model is the most widely used one-parameter logistic (1PL) form: under the 1PL, the probability that examinee i answers item j correctly depends only on the examinee's ability $\theta_i$ and the item's difficulty $b_j$:

$$P\left(X_{ij} = 1 \mid \theta_i, b_j\right) = \frac{exp\left(\theta_i - b_j\right)}{1 + exp\left(\theta_i - b_j\right)}$$

Higher $\theta$ indicates a more able examinee; higher $b$, a harder item. All 1PL fits in this study are performed with the mirt package (Chalmers, 2012).

The first level of evaluation (Figure 2c) asks whether the overall ability distribution of the simulated pool matches that of human examinees. We place all examinees on a common ability scale by anchoring. A 1PL model is first fit to the human-examinee data; the resulting abilities are standardized to mean 0 and SD 1, defining the reference scale. Each LLM × condition response matrix is then scored against the human-data item parameters and transformed by the same standardization, yielding ability estimates for the LLM examinees $\theta$ that are directly comparable to the human distribution. For each cell, we report the mean and SD of the LLM examinee $\theta$, together with the Overlapping Coefficient (OVL; Inman & Bradley, 1989) between the LLM examinee $\theta$ density and the human examinee $\theta$ density. Both densities are estimated with a Gaussian kernel and a common data-driven band width; OVL $\in [0, 1]$, with 1 indicating identical distributions. Larger OVL, and a mean and SD closer to 0 and 1, indicate better distribution-level alignment.

The second level analyzes behavior within profiles. As an initial, broader check, the simulated pool is divided into six subdistributions based on mastery counts (0–5 attributes mastered). These also act as a manipulation check: a well-steered LLM should show a steady increase in ability as the mastery count rises. We then compare the 32 distinct mastery profiles, asking whether examinees assigned a given profile attain the score expected of human examinees holding that profile. For each profile, we compute a human reference value from the CDM fit to the real data: the CDM-implied expected total score, the sum over the 15 items of the CDM-implied success probabilities for that profile (0–15). The corresponding LLM value is the observed mean total score over the C2 examinees assigned that profile. Because profiles differ greatly in how often they occur, each profile is weighted by its estimated pattern prevalence $\hat{P}(\alpha)$, and alignment is summarized by the prevalence-weighted Pearson correlation ($r_w$), the weighted signed bias (weighted mean of LLM minus human), and the weighted root-mean-square error (RMSE).

The third level compares simulated and human item difficulty on the two indices introduced above, the CTT pass rate and the 1PL difficulty parameter $b_j$. For each cell, we compute four diagnostics against the human-data values: the Spearman rank correlation ($\rho$) for relative item structure, the Pearson correlation ($r$) for absolute linear alignment, the signed bias (mean of LLM-derived minus human values), and RMSE for overall recovery error.

# 4 Results

## 4.1 Distribution-level alignment

Table 1 presents the anchored ability statistics. Under the no-profile baseline (C1), the simulated pools differed greatly from the human distribution in both location and spread. The means were shifted away from the human value of 0, often well above it (for example, +1.35 for Gemini 3.0 Flash (Thinking) and +1.40 for DeepSeek V3.2 (Thinking)), and the standard deviations were far too small, falling below 0.45 in seven of the eight cells (as low as 0.03 for Gemini 3.0 Flash (Non-Thinking)). The overlap with the human distribution was also low, with OVL ranging from 0.32 to 0.53. These two problems, a shifted mean and a narrow spread, are precisely the less-alignment and less-variability limitations that motivate CDP.

Adding profiles mitigated both problems for most configurations. Compared with C1, the uninformative condition (C2) raised OVL for every configuration and widened the standard deviation toward 1 in seven of the eight; the effect on the mean was mixed, moving it toward 0 for the configurations that started well above the human mean while pushing the already low means of the weaker configurations further below 0. The informative condition (C3) improved alignment further for seven of the eight configurations. OVL was highest under C3 (for Gemini 3.1 Pro it rose from 0.52 at C1 to 0.77 at C2 and 0.81 at C3; for Gemini 3.0 Flash (Thinking), from 0.36 to 0.81 to 0.83), the mean was generally closest to 0 (for example, +0.08 for Gemini 3.1 Pro and +0.16 for Gemini 3.0 Flash (Thinking)), and the standard deviation generally closest to 1. The one exception was DeepSeek V3.2 (Thinking), whose C2 already reached the highest overlap (0.84); adding the prior lowered it to 0.71. Among the two Gemini 3.0 Flash variants, the Thinking configuration aligned better than the Non-Thinking configuration (OVL of 0.81 and 0.83 versus 0.67 and 0.79 under C2 and C3), and the two thinking Gemini configurations reached the highest overlap under C3. Even under C3, however, overlap for the two Qwen configurations and Mistral Small 3.2 remained near 0.5, so profiling narrowed their gap without closing it.

**Table 1**

*Person-level ability alignment by configuration and condition*

| Configuration | Condition | Mean | SD | OVL |
|---|---|---|---|---|
| Gemini 3.1 Pro (Thinking) | C1 | 0.75 | 0.96 | 0.52 |
| | C2 | −0.56 | 0.87 | 0.77 |
| | C3 | 0.08 | 1.19 | 0.81 |
| Gemini 3.0 Flash (Thinking) | C1 | 1.35 | 0.22 | 0.36 |
| | C2 | −0.48 | 0.91 | 0.81 |
| | C3 | 0.16 | 1.16 | 0.83 |
| Gemini 3.0 Flash (Non-Thinking) | C1 | 0.74 | 0.03 | 0.35 |
| | C2 | −0.25 | 0.61 | 0.67 |
| | C3 | 0.11 | 0.80 | 0.79 |
| DeepSeek V3.2 (Thinking) | C1 | 1.40 | 0.17 | 0.32 |
| | C2 | 0.20 | 0.85 | 0.84 |
| | C3 | 0.60 | 0.86 | 0.71 |
| DeepSeek V3.2 (Non-Thinking) | C1 | 0.43 | 0.42 | 0.53 |
| | C2 | −0.57 | 0.52 | 0.56 |
| | C3 | −0.05 | 0.82 | 0.78 |
| Qwen 3.5 122B | C1 | 0.07 | 0.25 | 0.40 |
| | C2 | −0.42 | 0.36 | 0.51 |
| | C3 | −0.22 | 0.46 | 0.56 |
| Qwen 3.5 27B | C1 | −0.29 | 0.27 | 0.45 |
| | C2 | −0.62 | 0.31 | 0.48 |
| | C3 | −0.39 | 0.46 | 0.56 |
| Mistral Small 3.2 | C1 | −0.56 | 0.23 | 0.43 |
| | C2 | −0.91 | 0.37 | 0.50 |
| | C3 | −0.82 | 0.37 | 0.52 |
| Human examinees | — | 0.00 | 1.00 | — |

*Note.* $\theta$ is anchored to the human-examinee 1PL scale (Human Mean = 0, SD = 1). OVL = overlapping coefficient between each LLM examinee $\theta$ distribution and the human examinee $\theta$ distribution (1 = identical). C1 = no-profile baseline; C2 = uninformative CDP; C3 = informative CDP.

## 4.2 Profile-level alignment

Whereas the previous section considered the overall distribution, the profile level examines how examinees behave within each mastery profile. Figure 3 shows the LLM examinee $\theta$ distributions for two configurations, Gemini 3.1 Pro (Thinking) and Qwen 3.5 27B, separated by the number of mastered attributes. Under C1, each pool formed a single narrow peak. Under C2 and C3, the subdistributions for 0 through 5 mastered attributes separated and shifted upward in order, so the simulated abilities spread across a wider range. Profiles also multiplied the variety of complete response patterns: Gemini 3.1 Pro (Thinking) produced 5 distinct patterns under C1 but 133 under C2 and 56 under C3, against 267 among the 536 human examinees (Appendix C). The pattern prevalences changed which subdistributions dominated: uniform assignment (C2) placed most examinees at the middle counts (about 31% at each of counts 2 and 3), whereas the pattern prevalences used in C3 placed 33% of examinees at full mastery (count 5) and 19% at count 1, shifting the pool toward the higher end and closer to the human distribution. The two configurations differed in the range they could reach. For Gemini 3.1 Pro (Thinking), the C3 subdistribution means ran from −1.36 at count 1 to +1.46 at count 5, covering the low to high range, whereas for Qwen 3.5 27B they ran only from −0.93 to +0.09, so even fully mastering examinees reached only about the average human level.

**Figure 3**

*Anchored ability (θ) distributions for two example configurations (Gemini 3.1 Pro (Thinking) and Qwen 3.5 27B) under the three conditions*

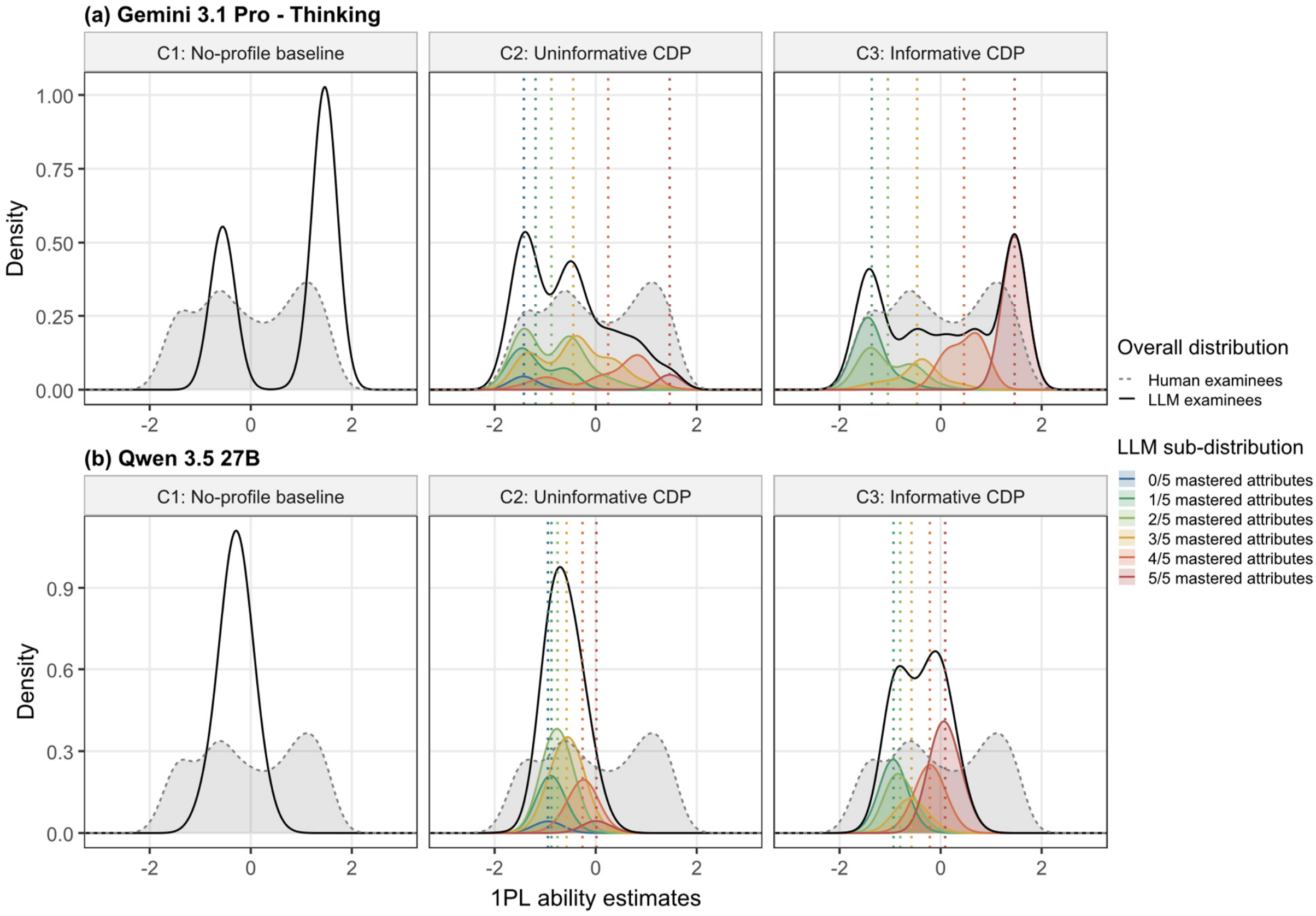

*Note.* Dashed line = human examinees; solid line = LLM examinees; shaded curves = LLM sub-distributions by number of mastered attributes. C1 = no-profile baseline; C2 = uninformative CDP; C3 = informative CDP.

Figure 4 plots the simulated profile means against the human expectation, and Table 2 reports the prevalence-weighted profile-level metrics. Across the 32 profiles, the simulated (C2) mean total scores closely matched the CDM-implied human expectation, with prevalence-weighted correlations $r_w$ between 0.92 and 0.98 for all eight configurations. The mastery-count aggregates in Figure 4 (triangles) rose steadily with the number of mastered attributes, showing that the LLMs responded to the assigned profiles as intended. Profile-level alignment is evaluated under C2 because its near-uniform allocation provides adequate examinee numbers for every profile, whereas under C3 low-prevalence profiles receive few or no examinees.

**Figure 4**

*Profile-level alignment between LLM-simulated (C2) and human examinees*

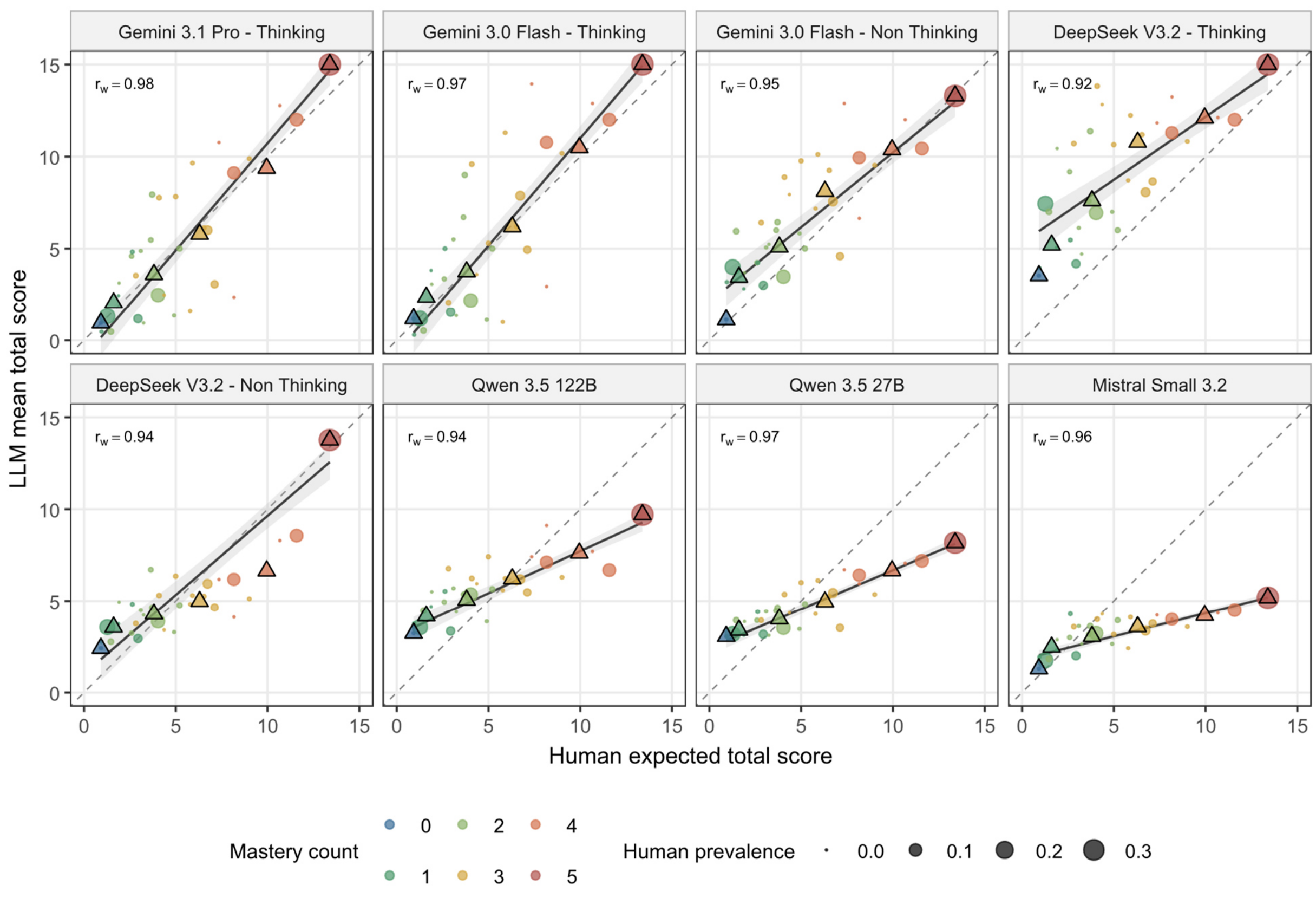

*Note.* Each point represents one of the 32 mastery profiles; the horizontal axis is the CDM-implied expected total score for the estimated human profile, and the vertical axis is the observed mean total score for examinees assigned to that profile. Point size is proportional to pattern prevalence; color denotes the number of mastered attributes; triangles mark the mastery-count aggregates; the gray line is the prevalence-weighted linear fit with its 95% confidence band; the dashed line is the identity. $r_w$ denotes the prevalence-weighted Pearson correlation.

**Table 2**

*Prevalence-weighted profile-level alignment between LLM (C2) and human examinees*

| Configuration | $r_w$ | $bias_w$ | $RMSE_w$ |
|---|---|---|---|
| Gemini 3.1 Pro (Thinking) | 0.98 | 0.40 | 1.54 |
| Gemini 3.0 Flash (Thinking) | 0.97 | 0.66 | 1.74 |
| Gemini 3.0 Flash (Non-Thinking) | 0.95 | 0.59 | 1.66 |
| DeepSeek V3.2 (Thinking) | 0.92 | 2.78 | 3.45 |
| DeepSeek V3.2 (Non-Thinking) | 0.94 | −0.09 | 1.61 |
| Qwen 3.5 122B | 0.94 | −1.22 | 2.94 |
| Qwen 3.5 27B | 0.97 | −2.19 | 3.54 |
| Mistral Small 3.2 | 0.96 | −4.20 | 5.51 |

*Note.* Prevalence-weighted alignment between LLM-simulated (C2) and human examinees across the 32 mastery profiles, with each profile weighted by its estimated pattern prevalence $\hat{P}(\alpha)$. The per-profile human value is

the CDM-implied expectation; the LLM value is the observed mean total score over examinees assigned that profile. $r_w$ = weighted Pearson correlation; $bias_w$ = weighted mean (LLM − human); $RMSE_w$ = weighted root-mean-square error. Total score = expected number correct (0–15).

The configurations differed mainly in absolute level. The weighted RMSE was smallest for Gemini 3.1 Pro (Thinking) ($bias_w$ = +0.40 and $RMSE_w$ = 1.54), followed by DeepSeek V3.2 (Non-Thinking) at 1.61 and the other two Gemini configurations at 1.66 and 1.74, whereas Mistral Small 3.2, Qwen 3.5 27B, and DeepSeek V3.2 (Thinking) departed most from the human expectation (weighted bias of −4.20, −2.19, and +2.78); DeepSeek V3.2 (Non-Thinking) showed the smallest bias (−0.09). Negative values indicate scores below the human expectation and positive values above it. This profile-level result suggests why the population information in C3 can be useful. When each profile already behaves correctly, the population's profile prevalences can combine these profiles into a pool that matches the human distribution and improves item-level recovery. The advantage of C3 over C2 is therefore consistent with a dependence on profile-level alignment, a point we return to in the Discussion.

## 4.3 Item-level recovery

Table 3 presents the recovery metrics for both difficulty indices, and Figures 5 and 6 plot the item-level comparisons for the CTT pass rate and the 1PL difficulty parameter. The baseline recovered item difficulty poorly. For Gemini 3.0 Flash (Thinking) and DeepSeek V3.2 (Thinking), C1 gave CTT pass-rate correlations near or below zero ($r$ = −0.06 and $r$ = −0.09, respectively), with large negative bias in the 1PL difficulty parameter (−5.7 to −6.0) and RMSE between 6.0 and 6.3, so items appeared far easier than they were. CDP improved recovery substantially. C2 raised the pass-rate and difficulty correlations for most configurations (for Gemini 3.0 Flash (Thinking) the CTT pass-rate correlation rose from −0.06 to 0.77, and for DeepSeek V3.2 (Thinking) from −0.09 to 0.72) and sharply reduced the bias and error in the 1PL difficulty parameter (its 1PL RMSE fell from about 6 to near 1 for both).

C3 improved absolute alignment further for most configurations, lowering the absolute bias and RMSE and raising the Spearman correlation on both indices (for Gemini 3.1 Pro, the RMSE for the CTT pass rate fell from 0.22 at C2 to 0.09 at C3, and the bias in the 1PL difficulty parameter moved from +1.16 to −0.20, with its RMSE falling from 1.40 to 0.73). A few cells were exceptions, where C2 matched or slightly beat C3 on absolute error, namely DeepSeek V3.2 (Thinking) on both indices, Gemini 3.0 Flash (Non-Thinking) on the 1PL difficulty index, and Mistral Small 3.2, which was about the same in both conditions. As at the distribution level, these exceptions are consistent with the role of profile-level alignment: the further item-level gain from the pattern prevalences tended to be larger when within-profile behavior was well aligned, and smaller or absent where that alignment was weaker. The thinking advantage was clearest at this level. Gemini 3.0 Flash (Thinking) did better than its Non-Thinking counterpart on both indices (the CTT pass-rate correlation was 0.80 versus 0.64 under C3, and the 1PL difficulty correlation was 0.76 versus 0.49 under C2), and DeepSeek V3.2 (Thinking) also did better than its Non-Thinking variant at the item level. The two thinking Gemini configurations again showed the strongest recovery, with Gemini 3.1 Pro (Thinking) reaching the best absolute alignment and Gemini 3.0 Flash (Thinking) showing the largest gain over its own baseline. Across the three levels the results were consistent: CDP improved on the baseline throughout, the informative condition added a further gain that depended on profile-level alignment, and the thinking Gemini configurations recovered both the ability distributions and item difficulty most faithfully, while the benefit of thinking for DeepSeek was confined to the item level.

**Table 3**

*Item-difficulty recovery (CTT and 1PL) by configuration and condition*

| Configuration | Condition | CTT difficulty | | | | 1PL difficulty | | | |
|---|---|---|---|---|---|---|---|---|---|
| | | r | ρ | bias | RMSE | r | ρ | bias | RMSE |
| Gemini 3.1 | C1 | 0.75 | 0.82 | 0.23 | 0.26 | 0.74 | 0.66 | −2.98 | 3.96 |
| Pro | C2 | 0.75 | 0.83 | −0.19 | 0.22 | 0.75 | 0.83 | 1.16 | 1.40 |
| (Thinking) | C3 | 0.81 | 0.89 | 0.02 | 0.09 | 0.75 | 0.89 | −0.20 | 0.73 |
| Gemini 3.0 | C1 | −0.06 | 0.33 | 0.44 | 0.47 | −0.06 | 0.24 | −5.96 | 6.31 |
| Flash | C2 | 0.77 | 0.87 | −0.16 | 0.20 | 0.76 | 0.86 | 1.00 | 1.30 |
| (Thinking) | C3 | 0.80 | 0.90 | 0.05 | 0.10 | 0.72 | 0.90 | −0.38 | 0.90 |
| Gemini 3.0 | C1 | 0.02 | 0.00 | 0.27 | 0.50 | 0.00 | 0.00 | −4.01 | 6.57 |
| Flash (Non- | C2 | 0.58 | 0.54 | −0.08 | 0.22 | 0.49 | 0.54 | 0.32 | 1.74 |
| Thinking) | C3 | 0.64 | 0.69 | 0.04 | 0.19 | 0.54 | 0.68 | −0.43 | 2.12 |
| DeepSeek | C1 | −0.09 | 0.36 | 0.45 | 0.48 | 0.17 | 0.08 | −5.72 | 5.95 |
| V3.2 | C2 | 0.72 | 0.70 | 0.07 | 0.16 | 0.75 | 0.70 | −0.51 | 1.07 |
| (Thinking) | C3 | 0.76 | 0.75 | 0.20 | 0.22 | 0.78 | 0.75 | −1.43 | 1.67 |
| DeepSeek | C1 | 0.36 | 0.37 | 0.16 | 0.35 | 0.49 | 0.37 | −2.29 | 3.80 |
| V3.2 (Non- | C2 | 0.50 | 0.40 | −0.20 | 0.36 | 0.49 | 0.40 | 1.52 | 2.70 |
| Thinking) | C3 | 0.49 | 0.45 | −0.02 | 0.25 | 0.47 | 0.45 | −0.06 | 1.99 |
| Qwen 3.5 | C1 | 0.47 | 0.50 | 0.03 | 0.37 | 0.50 | 0.50 | −0.78 | 4.20 |
| 122B | C2 | 0.65 | 0.47 | −0.15 | 0.34 | 0.59 | 0.47 | 0.84 | 3.65 |
| | C3 | 0.60 | 0.51 | −0.07 | 0.31 | 0.62 | 0.51 | 0.22 | 2.99 |
| Qwen 3.5 | C1 | 0.50 | 0.53 | −0.10 | 0.39 | 0.55 | 0.54 | 0.83 | 4.50 |
| 27B | C2 | 0.58 | 0.47 | −0.22 | 0.38 | 0.52 | 0.42 | 2.45 | 4.56 |
| | C3 | 0.58 | 0.48 | −0.13 | 0.34 | 0.53 | 0.48 | 1.19 | 3.78 |
| Mistral Small | C1 | 0.71 | 0.41 | −0.20 | 0.38 | 0.64 | 0.40 | 2.71 | 4.46 |
| 3.2 | C2 | 0.76 | 0.74 | −0.31 | 0.36 | 0.73 | 0.74 | 2.82 | 3.38 |
| | C3 | 0.75 | 0.73 | −0.28 | 0.35 | 0.73 | 0.73 | 2.79 | 3.50 |

*Note.* $r$ = Pearson correlation, $\rho$ = Spearman correlation between the LLM-derived and human-examinee item-difficulty indices; bias = mean(LLM − human); RMSE = root-mean-square error. CTT difficulty = item pass rate; 1PL difficulty = item-difficulty parameter $b$.

## Figure 5

*Recovery of classical test theory (CTT) item difficulty*

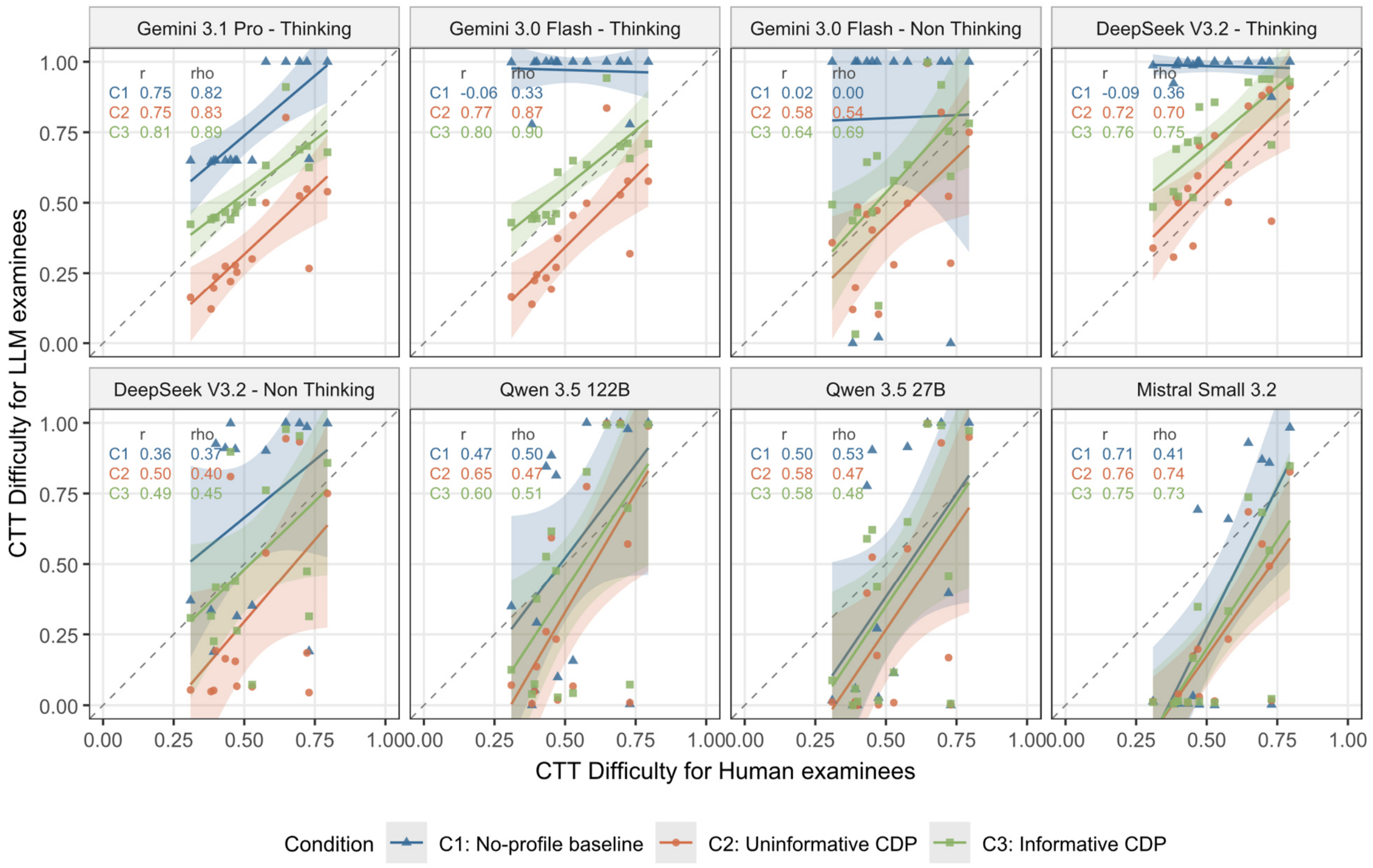

*Note.* Each of the 15 items compares CTT difficulty (pass rate) from LLM-simulated and human data for each LLM × condition cell; the dashed line is the identity, and colored lines are linear fits. $r$ = Pearson correlation, $\rho$ = Spearman correlation.

**Figure 6**

Recovery of IRT 1PL item difficulty ($b$)

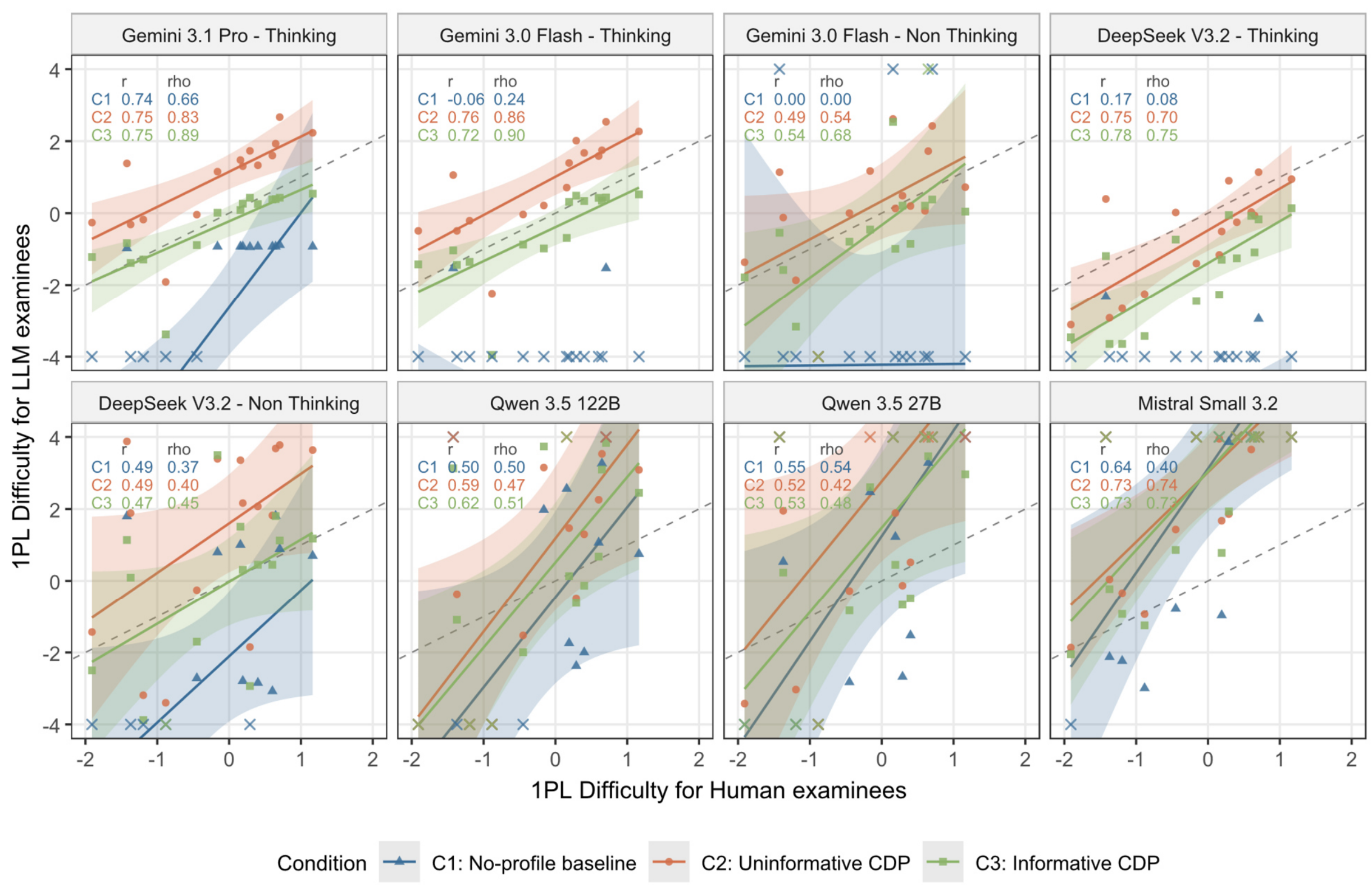

*Note.* Each of the 15 items compares 1PL difficulty from LLM-simulated and human data for each LLM × condition cell; the dashed line is the identity, and colored lines are linear fits. $r$ = Pearson correlation, $\rho$ = Spearman correlation.

# 5 Discussion and Conclusions

This study proposed Cognitive Diagnostic Profiling (CDP) and evaluated it against a human examinee reference across eight model configurations and three prompting conditions. CDP improved alignment at three levels: the overall ability distribution, the individual mastery profile, and item difficulty parameters, countering the less-alignment and less-variability failures observed in the no-profile baseline.

CDP increases response diversity through the cognitive structure of the test. Without profiles, even the strongest configuration produced only 5 distinct response patterns, with 64.4% being the all-correct pattern; with profiles, the same configuration produced 133 patterns. Each response traces back to an interpretable skill state, so the added variability directly reflects what the assessment measures. Decoding parameters such as temperature or top-K also increase variability but without connection to the construct. The informative condition, which sampled profiles according to their prevalence in the target population, brought further gains when simulated examinees already behaved consistently within each profile. Where within-profile behavior was less accurate, adding pattern prevalences brought little further gain. DeepSeek V3.2 (Thinking) illustrates this: its examinees overshot the human expectation within profiles (weighted bias +2.78), so a prior concentrating a third of the pool at full mastery amplified the overshoot, and overlap fell from 0.84 to 0.71. Faithful profile-level responding therefore matters more than the choice of sampling distribution. Model capability sets a further bound: even with full mastery assigned, Qwen 3.5 27B reached only about the average human level, while Gemini 3.1 Pro (Thinking) covered the full range.

Beyond these specific findings, the results carry broader implications. First, the results support a long-standing principle of assessment design: measurement should rest on an explicit model of cognition (National Research Council, 2001). Our findings extend this principle to simulation. Without a cognitive model in the prompt (C1), simulated examinees collapsed toward a single response pattern. Once the mastery profile made the model explicit (C2 and C3), alignment improved at every level. A practical corollary is that the artifacts assessment design already produces, such as student models (Mislevy et al., 2003), construct maps (Wilson, 2023), and the attribute specifications and Q-matrices of diagnostic assessment (Tatsuoka, 1983; Nichols, 1994; Leighton & Gierl, 2007), are exactly the inputs CDP needs. Testing programs that maintain these artifacts can adopt profile-based simulation without extra specification work. Second, the three-tier evaluation framework used here (distribution, profile, item) revealed discrepancies that no single indicator would have captured: a configuration can match the overall distribution while misaligning at the profile level, or rank items correctly while producing parameter values far from the human scale (Liu et al., 2025; Brant et al., 2026). We suggest that future studies evaluating LLM-based simulation adopt multi-level assessment (He-Yueya et al., 2024). Profile-level analysis, in particular, offers an actionable diagnostic: it pinpoints which mastery patterns the model handles well and where fidelity breaks down. Third, CDP can serve as a data-augmentation tool when human examinees are scarce or cluster in a narrow range of cognitive profiles (Liu et al., 2026; Ormerod, 2026). The cost ranges from less than 0.03 USD to about 14 USD per 100 simulated examinees depending on the configuration (Appendix B), orders of magnitude below human data collection; among the configurations tested, Gemini 3.0 Flash (Thinking) offered the best balance of cost and quality. Fourth, for educational technology platforms that rely on continually refreshed item pools, such as adaptive testing systems, online course platforms, and intelligent tutoring systems, CDP offers a scalable way to produce difficulty estimates for newly written items before any human data exist, narrowing a large pool to a smaller set worth field testing. The framework operates zero-shot and adapts to new tests through prompt changes alone, so it can be integrated into automated item-authoring pipelines. Profile-conditioned simulation also supports diagnostic assessment design: item developers can examine how simulated examinees with different mastery patterns respond and evaluate whether a new item discriminates among the intended cognitive profiles before any real student takes it. Fifth, profile-conditioned simulation can support fairness review. In operational testing, some mastery patterns are rare in the field-test sample, making it difficult to detect items that function differently for those examinees. By generating responses from the full range of profiles, CDP provides preliminary evidence on how items behave across cognitive subgroups, flagging potential issues before human field testing. In all cases, CDP complements human calibration.

In sum, CDP offers a structured, low-cost way to bring LLM-simulated examinees closer to human examinees across multiple psychometric dimensions, providing a fast first approximation that can make the early stages of item development more efficient.

# 6 Limitations and Future Work

The present findings provide initial evidence for CDP, and several limitations point to directions for future work. First, the evidence comes from a single, well-studied instrument: 15 fraction-subtraction items with five attributes and 536 examinees. Only one five-attribute decomposition was examined, though a single construct admits multiple decompositions that could yield different profiles and different alignment (de la Torre, 2008; Akbay et al., 2017). Even with CDP, the best configuration produced 133 distinct response patterns against 267 from human examinees (Appendix C), because attributes are a condensed summary of what the items demand and skills touched by only one or two items are usually dropped from the Q-matrix. Future studies should replicate CDP across content domains, attribute structures including hierarchically organized attributes (Köhn & Chiu, 2019), and longer tests, and explore whether defining attributes at finer granularity pushes simulated diversity closer to human levels. Second, the dataset is publicly available via the R package CDM and has been widely analyzed, so the items may appear in LLMs' training corpora. Profile conditioning drives the contrasts and C1 does not reproduce human statistics by default, so recall alone is unlikely to explain the effects, but contamination cannot be ruled out; replication on secure, unreleased item pools is needed to control for this possibility. Third, the informative condition used an in-sample prior: the pattern prevalences were estimated from the same 536 examinees that define the evaluation reference, so C3 marks an upper bound on what population information can contribute. Degraded-prior experiments—using priors drawn from earlier cohorts, expert judgment, or deliberately perturbed distributions—would quantify how quickly the C3 gain degrades with prior quality. Fourth, each cell was generated once at default sampling settings, leaving generation variability unquantified, and the results are tied to specific commercial model versions whose behavior and pricing change over time. Systematic comparisons across model families, scales, and reasoning settings would clarify which model properties drive alignment and how robust the advantage of reasoning-enabled models is. Fifth, CDP automated only the response-generation stage; the construct decomposition and profile rendering were authored manually. A multi-agent pipeline could automate the full workflow from construct decomposition through profile rendering to response generation. The generated response data also open a cross-validation opportunity: comparing response patterns against an expert-defined Q-matrix or an LLM-constructed Q-matrix can serve as a mutual consistency check, supporting Q-matrix validation and refinement (Sachdeva & Park, 2025; Du et al., 2026). Finally, the alignment examined here is psychometric rather than cognitive: matching distributions, profile expectations, and item parameters does not imply that the models solve items through human-like processes, and the manipulation check is a necessary rather than sufficient condition for treating simulated examinees as substitutes for human ones (Hullman et al., 2026). The gap between the semantic space of LLMs and the behavioral space of CDMs (Dong et al., 2025) is narrowed by CDP's natural-language rendering but not closed. Future work should analyze the chain-of-thought traces produced by reasoning models to examine whether the problem-solving steps align with the cognitive operations specified in the Q-matrix, moving the validation from outcome level to process level. Downstream validation should also examine whether item parameters calibrated from simulated examinees support their intended uses, for example through reliability and dimensionality checks on the resulting instruments. Hybrid designs that combine a small human sample with CDP-simulated examinees may offer a practical middle ground.

## Appendix A. Prompt Design

Each simulated examinee received one prompt. Under the no-profile baseline (C1), the prompt contained a role and task instruction, the 15 problems, and an output-format specification (A.1). Under the two profiling conditions, the prompt additionally included a target student profile block describing the examinee's mastery of the five skills, and a short instruction block (A.2). Conditions C2 (uninformative CDP) and C3 (informative CDP) used the identical prompt template and differed only in how each examinee's mastery pattern was sampled. The profile block tagged each of the five skills as MASTERED or NOT MASTERED with a fixed description of that state; any of the 32 mastery patterns was produced by selecting, for each skill, its mastered or not-mastered description. A.2 shows the full prompt for the all-mastered profile [1,1,1,1,1] and the profile block for the all-not-mastered profile [0,0,0,0,0], which together display both description variants for every skill.

### A.1 Baseline prompt (C1)

```
You are an experienced middle school math teacher. Predict the EXACT answer a
typical student would write for each fraction subtraction problem below.

PROBLEMS:
1. 3/4 - 3/8 = ?
2. 3 1/2 - 2 3/2 = ?
3. 6/7 - 4/7 = ?
4. 3 - 2 1/5 = ?
5. 3 7/8 - 2 = ?
6. 4 4/12 - 2 7/12 = ?
7. 4 1/3 - 2 4/3 = ?
8. 11/8 - 1/8 = ?
9. 3 4/5 - 3 2/5 = ?
10. 2 - 1/3 = ?
11. 4 5/7 - 1 4/7 = ?
12. 7 3/5 - 4/5 = ?
13. 4 1/10 - 2 8/10 = ?
14. 4 - 1 4/3 = ?
15. 4 1/3 - 1 5/3 = ?

OUTPUT FORMAT: Reply with JSON only, no explanations.
{"1":"answer","2":"answer","3":"answer","4":"answer","5":"answer","6":"answer","7
":"answer","8":"answer","9":"answer","10":"answer","11":"answer","12":"answer","1
3":"answer","14":"answer","15":"answer"}
```

### A.2 Profile prompt (C2 and C3)

Full prompt for the all-mastered profile [1,1,1,1,1]:

```
You are an experienced middle school math teacher. Predict the EXACT answer a
```

```
specific student would write for each fraction subtraction problem below.

TARGET STUDENT PROFILE [1,1,1,1,1]:

Skill 1 - Basic Fraction Subtraction Operation [MASTERED]:
    MASTERED: The student has mastered basic fraction subtraction operations.
They can correctly subtract fractions with same denominators and find common
denominators when they differ.

Skill 2 - Simplifying and Reducing Fractions [MASTERED]:
    MASTERED: The student has mastered simplifying and reducing fractions. They
can identify common factors between numerator and denominator and recognize when
results are improper fractions that need conversion to mixed numbers.

Skill 3 - Separating Whole Number from Fraction [MASTERED]:
    MASTERED: The student has mastered separating whole numbers from fractions in
mixed number operations. They understand the structural relationship between
whole numbers and fractions in mixed numbers and can flexibly separate,
recombine, or reorganize these parts during computation.

Skill 4 - Borrowing from Whole Number to Fraction [MASTERED]:
    MASTERED: The student has mastered borrowing from whole numbers to fractions.
They recognize when the minuend’s fraction is smaller than the subtrahend’s
fraction and correctly execute borrowing: decrementing the whole number by 1 and
adding the equivalent fraction to the fractional part.

Skill 5 - Converting Whole Number to Fraction [MASTERED]:
    MASTERED: The student has mastered converting whole numbers to equivalent
fractions. They know that n = n*d/d for any denominator d and apply this flexibly
when a pure whole number must participate in fraction subtraction.

PROBLEMS:
1. 3/4 - 3/8 = ?
2. 3 1/2 - 2 3/2 = ?
3. 6/7 - 4/7 = ?
4. 3 - 2 1/5 = ?
5. 3 7/8 - 2 = ?
6. 4 4/12 - 2 7/12 = ?
7. 4 1/3 - 2 4/3 = ?
8. 11/8 - 1/8 = ?
9. 3 4/5 - 3 2/5 = ?
10. 2 - 1/3 = ?
11. 4 5/7 - 1 4/7 = ?
12. 7 3/5 - 4/5 = ?
13. 4 1/10 - 2 8/10 = ?
14. 4 - 1 4/3 = ?
15. 4 1/3 - 1 5/3 = ?

INSTRUCTIONS:
- Even students with identical profiles show individual variation. Decide each
response based solely on the skill profile.
```

```
OUTPUT FORMAT: Reply with JSON only, no explanations.
{"1":"answer","2":"answer","3":"answer","4":"answer","5":"answer","6":"answer","7
":"answer","8":"answer","9":"answer","10":"answer","11":"answer","12":"answer","1
3":"answer","14":"answer","15":"answer"}
```

Profile block for the all-not-mastered profile [0,0,0,0,0] (the rest of the prompt is the same as above):

```
TARGET STUDENT PROFILE [0,0,0,0,0]:

Skill 1 - Basic Fraction Subtraction Operation [NOT MASTERED]:
    NON-MASTERED: The student has NOT mastered basic fraction subtraction
operations. They make systematic errors in the fundamental subtraction procedure
for fractions.

Skill 2 - Simplifying and Reducing Fractions [NOT MASTERED]:
    NON-MASTERED: The student has NOT mastered simplifying and reducing
fractions. They do not recognize when fractions need simplification or lack
procedural knowledge to find common factors.

Skill 3 - Separating Whole Number from Fraction [NOT MASTERED]:
    NON-MASTERED: The student has NOT mastered separating whole numbers from
fractions in mixed number operations. They do not understand how whole numbers
and fractions relate within mixed numbers, producing systematic structural
errors.

Skill 4 - Borrowing from Whole Number to Fraction [NOT MASTERED]:
    NON-MASTERED: The student has NOT mastered borrowing from whole numbers to
fractions. They cannot correctly execute borrowing in fraction subtraction,
producing systematic errors.

Skill 5 - Converting Whole Number to Fraction [NOT MASTERED]:
    NON-MASTERED: The student has NOT mastered converting whole numbers to
equivalent fractions. They cannot conceptualize whole numbers as fractions,
producing systematic errors when whole numbers must be converted.
```

# Appendix B. Cost and efficiency

Table A1 reports the average token usage and the resulting API cost for each configuration, expressed per 100 simulated examinees at May 2026 list prices. Prompt length was similar across configurations (about 0.065 to 0.069 million tokens per 100 examinees), so the cost differences came almost entirely from the reasoning tokens generated by the thinking configurations. The two thinking Gemini configurations produced more than 1 million reasoning tokens per 100 examinees, and DeepSeek V3.2 (Thinking) about 0.76 million, while the non-thinking configurations produced none.

The absolute cost remained low. The most expensive configuration, Gemini 3.1 Pro (Thinking), cost about 14.1 USD per 100 examinees, which is manageable for an item-calibration study. Gemini 3.0 Flash (Thinking) gave the best balance of cost and quality, nearly matching Gemini 3.1 Pro on alignment and recovery (Sections 4.1 to 4.3) at about 3.9 USD per 100 examinees, roughly a quarter of the cost. The remaining configurations cost less than 0.3 USD per 100 examinees, and the non-thinking open configurations less than 0.03 USD, so they remain worth considering once their lower alignment is weighed against the savings. By comparison, collecting the same 15-item instrument from human respondents on a crowdsourcing platform would cost far more (Prolific, 2026).

**Table A1**

*Token usage and API cost per 100 simulated examinees*

| Configuration | Prompt (M) | Output (M) | Reasoning (M) | Cost (USD) |
|---|---|---|---|---|
| Gemini 3.1 Pro (Thinking) | 0.067 | 0.015 | 1.149 | 14.108 |
| Gemini 3.0 Flash (Thinking) | 0.067 | 0.014 | 1.260 | 3.854 |
| Gemini 3.0 Flash (Non-Thinking) | 0.067 | 0.014 | 0.000 | 0.077 |
| DeepSeek V3.2 (Thinking) | 0.066 | 0.014 | 0.762 | 0.227 |
| DeepSeek V3.2 (Non-Thinking) | 0.065 | 0.013 | 0.000 | 0.013 |
| Qwen 3.5 122B | 0.069 | 0.012 | 0.000 | 0.029 |
| Qwen 3.5 27B | 0.069 | 0.013 | 0.000 | 0.025 |
| Mistral Small 3.2 | 0.068 | 0.022 | 0.000 | 0.013 |

## Appendix C. Response-pattern diversity

Table C1 summarizes the diversity of complete response patterns, the 15-item score vectors, for each configuration and condition, with the 536 human examinees as the reference. Distinct patterns count the unique vectors among the 536 simulated examinees; top pattern gives the share of the single most frequent vector; effective patterns is 2 raised to the Shannon entropy of the pattern distribution, the number of equally frequent patterns that would yield the same diversity.

**Table C1**

*Response-pattern diversity by configuration and condition*

| Configuration | Condition | Distinct patterns | Top pattern (%) | Effective patterns |
|---|---|---|---|---|
| Gemini 3.1 Pro (Thinking) | C1 | 5 | 64.4 | 2.1 |
| | C2 | 133 | 21.8 | 37.3 |
| | C3 | 56 | 33.0 | 11.4 |
| Gemini 3.0 Flash (Thinking) | C1 | 3 | 77.8 | 1.7 |
| | C2 | 52 | 22.2 | 20.6 |
| | C3 | 33 | 33.0 | 10.1 |
| Gemini 3.0 Flash (Non-Thinking) | C1 | 2 | 98.0 | 1.1 |
| | C2 | 103 | 6.7 | 62.2 |
| | C3 | 55 | 25.2 | 19.8 |
| DeepSeek V3.2 (Thinking) | C1 | 9 | 86.9 | 1.8 |
| | C2 | 104 | 9.9 | 44.5 |
| | C3 | 75 | 36.8 | 17.7 |
| DeepSeek V3.2 (Non-Thinking) | C1 | 75 | 17.4 | 37.7 |
| | C2 | 98 | 24.1 | 24.4 |
| | C3 | 94 | 20.0 | 26.2 |
| Qwen 3.5 122B | C1 | 74 | 19.4 | 26.5 |
| | C2 | 89 | 13.6 | 28.5 |
| | C3 | 92 | 9.9 | 39.0 |
| Qwen 3.5 27B | C1 | 38 | 27.2 | 13.6 |
| | C2 | 45 | 21.5 | 18.5 |
| | C3 | 48 | 22.0 | 19.9 |
| Mistral Small 3.2 | C1 | 56 | 35.6 | 12.4 |
| | C2 | 119 | 9.9 | 50.3 |
| | C3 | 101 | 9.1 | 46.4 |
| Human examinees | — | 267 | 10.3 | 135.3 |